\documentclass[structabstract]{aa}

\usepackage{graphicx}
\usepackage{txfonts}
\usepackage{natbib}
\usepackage{lscape}
\usepackage{float}
\usepackage{placeins}
\bibpunct{(}{)}{;}{a}{}{,}

\begin{document}
\title{The Australia Telescope Large Area Survey:\\2.3\,GHz observations of ELAIS-S1 and CDF-S}

\subtitle{Spectral index properties of the faint radio sky}

\author{Peter-Christian Zinn\inst{1,2}
  \and
  Enno Middelberg\inst{1}
  \and
  Ray P. Norris\inst{2}
  \and
  Christopher A. Hales\inst{3,2}
  \and
  Minnie Y. Mao\inst{2,4,5}
  \and
  Kate E. Randall\inst{2,3}
}

\institute{Astronomical Institute of Ruhr-University Bochum, Universit\"atsstra\ss{}e 150, 44801 Bochum, Germany\\
  \email{zinn@astro.rub.de}
\and
CSIRO Astronomy \& Space Science, PO Box 76, Epping, NSW, 1710, Australia
\and
Sydney Institute for Astronomy, School of Physics, The University of Sydney, NSW 2006, Australia
\and
Australian Astronomical Observatory, PO Box 296, Epping, NSW 1710, Australia
\and
School of Mathematics \& Physics, University of Tasmania, Private Bag 37, Hobart, 7001, Australia
}

\date{Received 04/05/2012; accepted 22/06/2012}

\abstract 
{The Australia Telescope Large Area Survey (ATLAS) aims to image a
  7\,deg$^2$ region centred on the European Large Area ISO Survey -
  South 1 (ELAIS-S1) field and the {\it Chandra} {\rm Deep Field South
    (CDF-S) at 1.4\,GHz with high sensitivity (up to
    $\sigma\sim10\,\mu$Jy) to study the evolution of star-forming
    galaxies (SFGs) and Active Galactic Nuclei (AGN) over a wide range
    of cosmic time.}}
{{\rm We present here ancillary radio observations at a frequency of
    2.3\,GHz obtained with the Australia Telescope Compact Array
    (ATCA). The main goal of this is to study the radio spectra of an
    unprecedented large sample of sources {($\sim2000$ observed, $\sim600$ detected in both frequencies)}.}}
{{\rm With this paper, we provide 2.3\,GHz source catalogues for both
    ATLAS fields, with a detection limit of 300\,$\mu$Jy (equivalent
    to $4.5\sigma$ in the ELAIS-S1 field and $4.0\sigma$ in the
    CDF-S). We compute spectral indices between 1.4\,GHz and 2.3\,GHz
    using matched-resolution images and investigate various properties
    of our source sample in dependence of their spectral indices.}}
{{\rm We find the entire source sample to have a median spectral index
    $\alpha_{\mathrm{med}}=-0.74$, in good agreement with both the
    canonical value of $-0.7$ for optically thin synchrotron radiation and other
    spectral index studies conducted by various groups. Regarding the
    radio spectral index as indicator for source type, we
    find only marginal correlations so that flat or inverted spectrum sources are usually powered by AGN and hence conclude
    that at least for the faint population the spectral index is not a
    strong discriminator. We investigate the $z$--$\alpha$ relation for our source sample and find no such correlation between spectral index and redshift at all. We do find a significant correlation between redshift and radio to near-infrared flux ratio, making this a  much stronger tracer of high-$z$ radio sources. We also find no evidence for a dependence of
    the radio--IR correlation on spectral index.}}
{}


\titlerunning{ATLAS 2.3\,GHz observations}
\maketitle

\section{Introduction}
\label{intro}

Large-scale radio surveys are the primary instrument in the toolbox of
contemporary astronomy to investigate demography and evolution of
active galaxies across cosmic time. Consequently, there have been
numerous attempts to use this tool, whether in a wide, shallow layout
\citep[e.g. the National Radio Astronomy Observatory Very Large Array Sky Survey,][]{Condon1998} or in a deep, but more
pencil beam-like layout such as for the Cosmic Evolution Survey
\citep[COSMOS,][]{Schinnerer2010} or even the Lockman Hole survey by
\cite{Owen2008}, the most sensitive radio observation to date.

The Australia Telescope Large Area Survey
\citep[ATLAS,][]{Norris2006,Middelberg2008} attempts to break this
dichotomy by obtaining deep images ($\sigma\,\sim\,10\,\mu$Jy in the
final data release 3) over an unprecedented large area of
sky. Distributed over the well-studied {\it Chandra} Deep Field South
(CDF-S) and the European Large Area ISO Survey -- South 1 (ELAIS-S1)
field, the total area covered by ATLAS is approximately seven square
degrees. This allows { us} to study the faint radio sky without being
affected by cosmic variance \citep{Moster2011}. Furthermore, the
wealth of ancillary photometric data ranging across the entire
electromagnetic spectrum \citep[e.g. from SWIRE, COMBO-17 or
  GEMS][]{Lonsdale2003,Wolf2003,Rix2004} as well as numerous
spectroscopic campaigns
\citep{Sacchi2009,Popesso2009,Balestra2010,Mao2012} enables one to
conduct detailed studies of radio-detected objects.

In this paper, we focus on radio spectral properties of the ATLAS
source sample. The main ATLAS data releases are DR\,1 by
\cite{Norris2006} and \cite{Middelberg2008}, DR\,2 by
Hales et al. (in prep.), and DR\,3 by Banfield et al. (in prep.). These data
consist { of} 1.4\,GHz observations as for most contemporary radio surveys,
and we here supplement these data with observations at 2.3\,GHz. The
main goal of this is to calculate spectral indices
$\alpha\equiv\log\left(S_{\mathrm{1.4\,GHz}}/S_{\mathrm{2.3\,GHz}}\right)/\log\left(1.4\,\mathrm{GHz}/2.3\,\mathrm{GHz}\right)$
for all sources detected at both wavelength and so to characterise
their radio emission mechanism, source type and dependence on other
spectral quantities. Even though many blind surveys of extragalactic
radio sources are available already, only few studies have been
carried out which characterise the radio spectra of a large source
population by means of their spectral index. For example,
\cite{Bondi2007} and \cite{Ibar2009} use supplementary 610\,MHz data
to calculate spectral indices at frequencies below 1.4\,GHz with the
main aim to identify steep spectrum sources which are regarded as good
candidates for being high-redshift objects (see
Sect. \ref{z-alpha}). The most comprehensive study of spectral indices
between 1.4\,GHz and a higher frequency was conducted by
\cite{Prandoni2006} using 5\,GHz Very Large Array (VLA) data in the
Australia Telescope ESO Slice Project survey field. With a sensitivity
of $\sim70\,\mu$Jy over an area of one square degree, their source
sample consists of approximately 100 objects which are detected at
both frequencies, so our sample exceeds this by a factor of six.

According to \cite{Komatsu2011}, we adopt a flat $\Lambda$CDM
cosmology with
$H_{\mathrm{0}}=70.2\,\mathrm{km}\,\mathrm{s}^{-1}\,\mathrm{Mpc}^{-1}$
and $\Omega_{\Lambda}=0.725$ throughout this paper.

\section{Observations and imaging}
\label{observations}
In the following description, we give the actual numbers for all
things concerning the ELAIS-S1 field first, followed by the numbers
for the CDF-S in parentheses. If there is only a single number given,
it is approximately equal for both fields. Observations at 2.3\,GHz
for the ATLAS fields were carried out between December 2006 and March
2007 on 52 separate days where the ATCA was only in its 750B and 750C
configurations. This resulted in a substantially lower resolution than
at 1.4\,GHz since the longest baselines used for imaging at 2.3\,GHz
was 750\,m, or 5.7\,k$\lambda$, resulting in an angular resolution of
$33\arcsec$ ($57\arcsec$), compared to
30\,k$\lambda$ at 1.4\,GHz. The ATCA primary beam at a frequency of
2.3\,GHz has a FWHM of 21\,arcmin, therefore 76 (88) overlapping
pointings were observed in a hexagonal lattice with an average
integration time of 6.17\,h (4.13\,h) per pointing (compared to
approx. 12\,h per pointing for the 1.4\,GHz image) to cover the entire
area of the 1.4\,GHz maps.

For phase calibration, which was done approximately every 45\,min,
PKS\,0022-423 (PKS\,0237-233) was observed as secondary
calibrator. Flux calibration was done for both fields using PKS
1943-638 which was observed at least once a day for 10\,min. As for
1.4\,GHz, the ATCA provides two independent frequency bands with a
bandwidth of 128\,MHz divided into 33 channels. During the first two
days of observations for the ELAIS-S1 field, the two bands were
centred at 2.368\,GHz and 2.560\,GHz in an attempt to reduce local
Radio Frequency Interference (RFI). The fluxes density of the primary
calibrator, PKS\,1934-638, was assumed to be 11.589\,Jy and
10.932\,Jy, respectively. Because using a higher frequency for the
upper band did not significantly decrease the RFI, the remaining time
was observed with the standard band configuration, meaning that the
upper band was lowered in central frequency to 2.496\,GHz. There, the
primary calibrator was assigned to have a flux density of
11.138\,Jy. We note that the final images were made with the entire
data set.

\begin{figure*}
\centering
\includegraphics[width=1.0\textwidth]{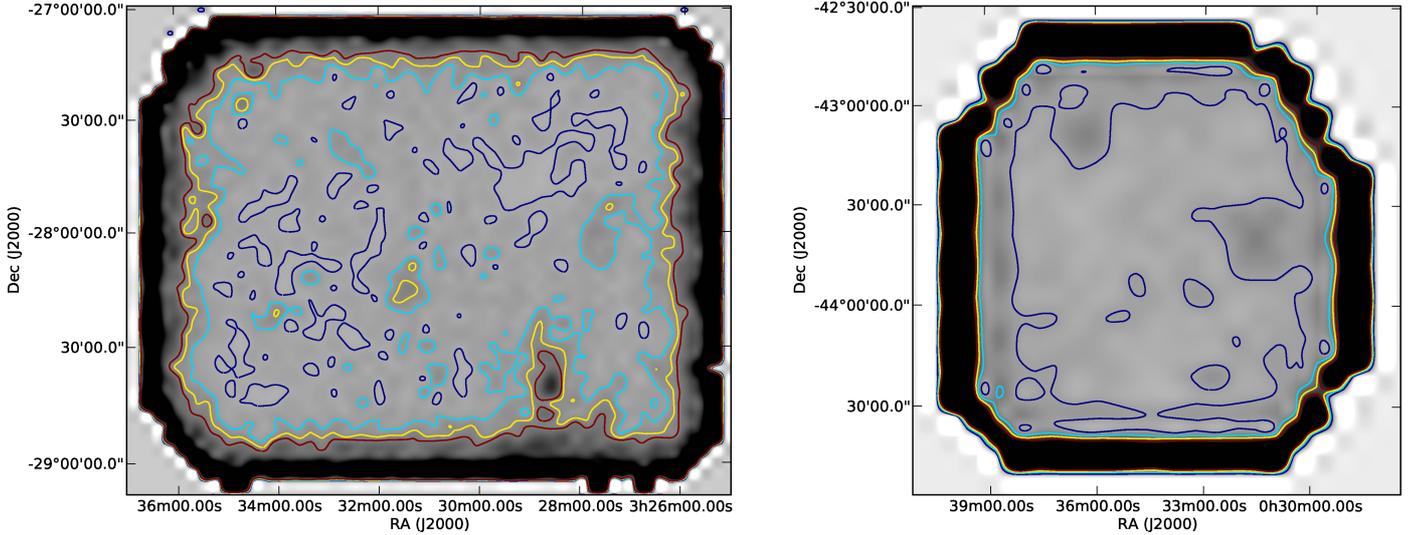}
\caption{2.3\,GHz noise maps of both ATLAS fields: CDF-S (left) and ELAIS-S1 (right). Contours start at 70\,$\mu$Jy\,beam$^{-1}$ (dark blue contour line) and increase by a constant value of 40\,$\mu$Jy\,beam$^{-1}$.}
\label{noisemap}
\end{figure*}

The calibration of the data were carried out using standard
\texttt{Miriad} \citep{Sault1995} tasks. \texttt{ATLOD} was used to
convert the raw data from \texttt{RPFITS} format to the native
\texttt{Miriad} format. For the further calibration, both channels at
the end of the bands were discarded due to a significant lack of
sensitivity, hence a data set containing two times 13 channels with
8\,MHz bandwidth each was obtained, yielding a net bandwidth of
208\,MHz. Fortunately, our data do not suffer from self-interference
typical for the ATCA. These so-called ``birdies'' occur within the
samplers at integer multiples of 128\,MHz. The 19$^{th}$ harmonic at
2.432\,GHz fortunately lies exactly between our two bands and
therefore does not affect our data. After bandpass-calibration, RFI
was removed using \texttt{Pieflag} \citep{Middelberg2006}, a tool that
compares statistics from channels with no or little RFI contamination
to the other channels in the data and looks for outliers and sections
with high noise. Bandpass calibration was carried out using the
observations of PKS\,1934-638. Phase calibration was done using the
regular scans of the secondary calibrators, and then amplitude
calibration was carried out using PKS\,1934-638. The data were
split into the individual pointings which were imaged separately using
uniform weighting. This provides a higher resolution at the cost of
lower sensitivity than natural weighting. The cell size was set to
$6\arcsec$ ($3\arcsec$). Because of the low fractional bandwidth
($<$\,10\%), a normal \texttt{CLEAN} procedure was used (in contrast
to the \texttt{Miriad} multi-frequency clean implementation
\texttt{MFCLEAN} for the 1.4\,GHz data) for deconvolution. Also the
dynamic range in the pointings was sufficiently low as to not require
any special deconvolution procedures. Cleaning was carried out using
2000 iterations, after which only marginal sidelobes around the two
strongest sources remained, and the effects of clean bias should be
negligible. After cleaning, the models were convolved with a restoring
beam of $33.56\arcsec\,\times\,19.90\arcsec$
($57.15\arcsec\,\times\,22.68\arcsec$) at a position angle of
$-1.32\,\deg$ ($-1.90\,\deg$). The final mosaics were produced using
\texttt{Miriad}'s \texttt{LINMOS} task to account for primary beam
attenuation and overlapping regions of the individual pointings. As a
result, the edges of the final images exhibit a significantly higher
noise level than the central regions (see Fig.~\ref{noisemap}).

\begin{table}
  \label{images}
  \caption{Parameters of the observations. Given are the number of
  pointings per field, the average integration time per pointing, the
  total area covered by the pointings, the rms noise and the
  resolution of the final images.}
  \centering
  \begin{tabular}{lccccc}
    \hline
    \hline
    \noalign{\smallskip}
    Field & N$_{\rm pnt}$ & t$_{\rm int} $ & area     & rms noise    & resolution\\
          &   & h      &    $\deg^2$ & $\mu$Jy/beam & arcsec\\
    \noalign{\smallskip}
    \hline
    \noalign{\smallskip}
     ELAIS-S1 & 76 & 6.17 & 2.77 & 70 & $33.56 \times19.90$\\
     CDF-S    & 88 & 4.13 & 3.63 & 80 & $57.15 \times22.68$\\
    \noalign{\smallskip}
    \hline
  \end{tabular}
  \tablefoot{{ For comparison, we note that the corresponding typical rms noise levels of the 1.4\,GHz images are 31\,$\mu$Jy and for 38\,$\mu$Jy the ELAIS-S1 and CDF-S, respectively.}}
\end{table}

\section{The source catalogue}
\label{catalog}
Source extraction from the final 2.3\,GHz images was performed in two
steps: First, all known 1.4\,GHz sources from ATLAS DR\,2
(Hales et al. in prep.) were used as an input catalogue to search for 2.3\,GHz
counterparts. { This catalog contains a total of 2240 individual 1.4\,GHz sources across both the ELAIS-S1 and the CDF-S of which 631 could be reliably detected at 2.3\,GHz.} To identify these counterparts, the 2.3\,GHz image was
inspected by eye at the location of every individual 1.4\,GHz
source. This was done by half-automatically browsing through an input
catalogue, simultaneously displaying the corresponding regions in the
1.4\,GHz and 2.3\,GHz images and, if a 2.3\,GHz counterpart was found,
fitting a Gaussian to it and recording all fit parameters in an output
catalogue. Thus, no blind search was conducted at 2.3\,GHz, allowing
the inclusion of sources fainter than the typical $5\,\sigma$ cutoff
in the catalogue. The lowest signal-to-noise ratio (SNR) among
  the 2.3\,GHz sources is 3.9. 

In a second step, we checked for ``2.3\,GHz-only'' sources. Those
sources are unlikely to occur because of the noise levels of the
1.4\,GHz and 2.3\,GHz images. Since the faintest 1.4\,GHz sources in
the DR\,2 catalogue are as faint as 100\,$\mu$Jy, whereas the faintest
sources in 2.3\,GHz are three times brighter, a 2.3\,GHz-only source
would have a highly inverted spectral index of
$\alpha\geq2.4$. Also, we note that, due to the temporal offsets between the 1.4\,GHz and 2.3\,GHz observations, possible variability of some radio sources may lead to flawed spectral indices. Nevertheless, we used \texttt{MIRIAD}'s \texttt{SAD}
task to automatically produce a 2.3\,GHz catalogue with a detection
threshold of $5\sigma$ and matched the sources obtained that way to
the 2.3\,GHz catalogue obtained by cross-matching to 1.4\,GHz
sources. All sources identified by \texttt{SAD} could be matched to
sources previously found as 1.4\,GHz counterparts, hence not a single
2.3\,GHz-only source was found.

To compute spectral indices, the flux densities in 1.4\,GHz had to be
re-measured because the resolutions of the 1.4\,GHz and 2.3\,GHz
images differ significantly. Whilst the 1.4\,GHz observations included
long baseline configurations of the ATCA (e.g. the 6A and 6B
configurations), the 2.3\,GHz observations were obtained with the ATCA
in 750B and 750C configurations only. This leads to a resolution three
to five times lower at 2.3\,GHz compared to the $\sim10\arcsec$
resolution of the 1.4\,GHz observations, despite the shorter
wavelength. Therefore, in order to avoid resolution effects, we
convolved the original 1.4\,GHz images with a Gaussian kernel to match
the resolution of the 2.3\,GHz images as given in Table
\ref{images}. We then used the ATLAS DR\,2 catalogue again as input to
re-measure the 1.4\,GHz flux densities in the low-resolution
images. Because the image sensitivity was reduced by a factor of
$\sim2$ by this convolution (which is equal to tapering the data in
the $uv$-plane, so in fact weighting down long baselines), all sources in the low-resolution image should be a true subset of the hi-res sources and, therefore, all low-resolution sources should be detected when using the high-resolution catalog as input sample.

During the re-measurement of the 1.4\,GHz flux densities using
convolved images, a problem surfaced with the ELAIS-S1 image. Unlike
with the CDF-S convolved 1.4\,GHz image, where the procedure outlined
above went well, the \texttt{MIRIAD} task \texttt{imfit} overestimated
the flux densities by a significant amount (up to 50\%). This became
obvious by inspecting residual images for sources with various flux
densities and degrees of compactness. While the residuals looked good
in the CDF-S, they showed bowls of negative flux in the ELAIS-S1
image. The cause of the effect remains unclear. The fitting routines
in the software package \texttt{CASA} \citep{CASA} yielded similarly
overestimated flux densities, so it appears that the effect is not
caused by the fitting software, but is inherent in the images, and we
had to work around this issue. This was done by iteratively fitting
Gaussians to every individual source. The first Gaussian fit (which
was always overestimating the flux density) was used as prior. The
resulting residual image was then examined in terms of total flux and
rms noise. If the total flux in the residual image (trimmed to a size
twice as large as the restoring beam) was negative and the rms noise
was deviating from the median rms in the region of the source by more
than 10\%, the fit was discarded and replaced by a fit with a peak
flux density 2\% lower than in the previous fit. The residual image
was again examined in a similar way. This procedure was iteratively
repeated until the criteria (non-negative total flux and rms noise
consistent with the rms in this region) were entirely met. As a last
step, all final residual images were inspected by eye to assure the
quality of the fit.

A further problem in calculating spectral indices for every 1.4\,GHz
was introduced by the very different resolutions of the 1.4\,GHz and
2.3\,GHz observations. Because the resolutions differ by a large
factor, as explained above, numerous sources which are clearly
separated in the high-res images are blended together in the low-res
images. Therefore, we chose to completely exclude such blended sources
from our final catalogue if they could not be separated by
simultaneously fitting two or more Gaussians. We note that this may
bias our sample towards fainter sources since classic radio doubles
(which are mainly excluded by the blending) are generally brighter
than single, isolated sources.

The final catalogue was then constructed by merging the individual
catalogues from ELAIS-S1 and CDF-S. This resulted in a final catalogue
comprising 631 radio sources with clear detections and well-measured
flux densities in both 1.4\,GHz and 2.3\,GHz. A portion of the catalogue
is shown in Table~\ref{catalogue} for illustration.

\begin{landscape} 
\begin{table}
  \caption{A representative portion of the ATLAS 2.3\,GHz source
    catalogue of the ELAIS-S1 field for illustration purposes. The
    CDF-S catalogue contains the same information.}
  \label{catalogue}
  \centering
  {
  \begin{tabular}{lrrrrrrrrrrrrrc}
    \hline
    \hline
    \noalign{\smallskip}
    ID\tablefootmark{a}&RA (J2000)&DEC (J2000)&$S_{\mathrm{1.4\,GHz,hi}}$&$\Delta S_{\mathrm{1.4\,GHz,hi}}$&$S_{\mathrm{1.4\,GHz,lo}}$&$\Delta S_{\mathrm{1.4\,GHz,lo}}$&$rms_{\mathrm{1.4\,GHz,lo}}$&$S_{\mathrm{2.3\,GHz}}$&$\Delta S_{\mathrm{2.3\,GHz}}$&rms$_{\mathrm{2.3\,GHz}}$&$\alpha^{2.3}_{1.4}$&$\Delta\alpha^{2.3}_{1.4}$&$z$&class\tablefootmark{b}\\
    &deg&deg&mJy&mJy&mJy&mJy&mJy&mJy&mJy&mJy&&&&\\
    \noalign{\smallskip}
    \hline
    \noalign{\smallskip}
254&51.770792&-28.650297&1.260&0.084&1.230&0.093&0.069&0.561&0.080&0.075&-1.58&0.11&&\\
178&51.621458&-27.427269&2.740&0.145&2.220&0.165&0.122&1.519&0.107&0.075&-0.76&0.11&&\\
165&53.319708&-27.941922&1.131&0.054&0.762&0.158&0.153&0.828&0.084&0.073&0.17&0.13&0.685&\\
163&52.470667&-27.492661&2.160&0.116&2.070&0.985&0.980&1.810&0.116&0.073&-0.27&0.95&&\\
102&52.561167&-28.566331&0.194&0.037&0.162&0.155&0.154&0.344&0.074&0.072&1.52&0.64&&\\
290&53.190750&-28.520756&0.714&0.044&0.598&0.126&0.123&0.400&0.075&0.072&-0.81&0.14&0.211&\\
201&52.399667&-27.950881&1.420&0.087&1.550&0.153&0.132&1.154&0.093&0.072&-0.59&0.06&0.067&SF\\
150&52.122000&-28.030825&1.810&0.087&1.770&0.281&0.267&1.016&0.088&0.072&-1.12&0.30&0.902&\\
264&52.105750&-27.745814&0.920&0.054&0.745&0.176&0.173&0.674&0.080&0.072&-0.20&0.46&&\\
323&52.808292&-28.785917&0.834&0.057&0.818&0.160&0.155&0.544&0.077&0.072&-0.82&0.38&0.140&AGNe\\
    \noalign{\smallskip}
    \hline
  \end{tabular}}
  \tablefoot{The columns are as follows: (1) unique ID (2) degrees of J2000 right ascension (3) degrees of J2000 declination (4,5) 1.4\,GHz integrated flux density and flux density error as given in Hales et al. (in prep) (6,7,8) integrated flux density, flux density error as calculated using Eqn.~\ref{error} and local rms noise as measured from the low-resolution 1.4\,GHz image (9,10,11) integrated flux density, flux density error as calculated using Eqn.~\ref{error} and local rms noise as measured from the 2.3\,GHz image (12,13) spectral index calculated from the low-resolution 1.4\,GHz flux density and the 2.3\,GHz flux density and its error as calculated by Eqn.~\ref{spixerr} (14,15) spectroscopic redshift and spectral classification from \cite{Mao2012} if available. The entire table is available via the on-line version of the journal.\\ \tablefoottext{a}{Unique only for the respective field
      (ELAIS-S1 or CDF-S).}  \tablefoottext{b}{Spectroscopic
      classification from \cite{Mao2012}. SF indicates an optical
      spectrum dominated by star formation activity while AGN
      indicates an optical spectrum typical for AGN (e.g. AGNe for an
      AGN with characteristic emission lines). E indicates that the
      spectrum is that of an early-type galaxy and hence the radio
      emission must originate from an AGN hosted by it.}}
\end{table}
\end{landscape}

We give an identifier for each source, the position, the original
1.4\,GHz flux density as measured by Hales et al. (in prep.), the re-measured
1.4\,GHz flux density using the convolved image, the 2.3\,GHz flux
density as measured during the cross-matching process and finally the
spectral index calculated from the matched-resolution 1.4\,GHz and
2.3\,GHz flux densities. All flux densities are integrated quantities.

Errors for the flux densities were either taken from Hales et al. (in prep.)
in case of the original 1.4\,GHz flux densities or calculated
following \cite{Hopkins2003} for the low-resolution 1.4\,GHz flux
densities and the 2.3\,GHz flux densities, taking into account both
the local rms noise at the source position as well as the quality of
the Gaussian fit:

\begin{equation}
\frac{\Delta S}{S}=\sqrt{\frac{\mu_{image}^2}{S^2}+\frac{\mu_{fit}^2}{S^2}},
\label{error}
\end{equation}

where $\mu_{image}$ quantifies the uncertainty in flux density due to
the image noise and $\mu_{fit}$ represents uncertainties due to the
fitting of a Gaussian. These two quantities can be calculated
following \cite{Windhorst1984},

\begin{equation}
\frac{\mu_{image}}{S}=\sqrt{\frac{\sigma^2}{S_{peak}^2}+C_f^2+C_p^2},
\label{rmsError}
\end{equation}

where $\sigma$ is the local rms noise, $C_f$ the error in absolute
flux calibration and $C_p$ the error in flux due to eventual pointing
errors. We here adopt a conservative value of
$C_f^2+C_p^2=0.05^2$. The fitting error is calculated following
\cite{Condon1997},

\begin{equation}
\frac{\mu_{fit}}{S}=\sqrt{\frac{\mu_S^2}{S_{peak}^2}+\left(\frac{\Theta_B\Theta_b}{\Theta_M\Theta_m}\right)\left(\frac{\mu_M^2}{\Theta_M^2}+\frac{\mu_m^2}{\Theta_m^2}\right)}.
\end{equation}

Here, $\Theta_B\Theta_b$ are the restoring beam's major and minor
axes, whereas $\Theta_M\Theta_m$ represent the major and minor axes of
the fitted Gaussian. The errors $\mu_S$,$\mu_M$ and $\mu_m$ stand for
the errors of the fitted Gaussian's peak flux, major and minor
axes. The error of the spectral index was finally calculated using
simple Gaussian error propagation,

\begin{equation}
\Delta\alpha=2\,\sqrt{\left(\frac{\Delta S_{1.4\,\rm{GHz}}}{S_{1.4\,\rm{GHz}}}\right)^2+\left(\frac{\Delta S_{2.3\,\rm{GHz}}}{S_{2.3\,\rm{GHz}}}\right)^2},
\label{spixerr}
\end{equation}

where the factor of $2\approx|[\log(1.4\,\rm{GHz}/2.3\,\rm{GHz})\cdot
  \ln(10)]^{-1}|$ arises from deriving the spectral index as defined
in section \ref{intro} with respect to the flux densities, which are
the only quantities with uncertainties in this equation (the error in
frequency is neglected).

\begin{figure*}
\centering
\includegraphics[width=0.49\textwidth]{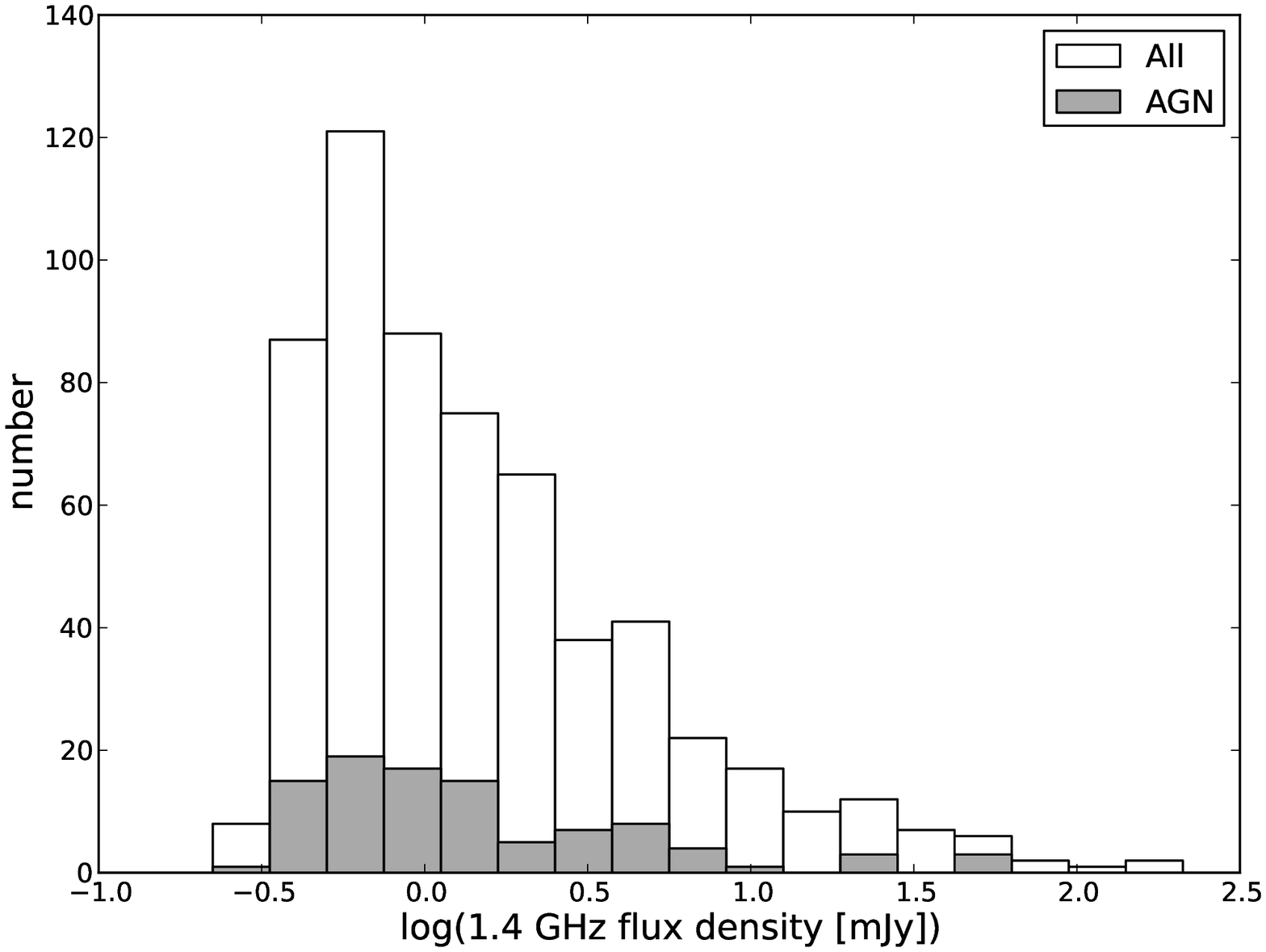}
\includegraphics[width=0.49\textwidth]{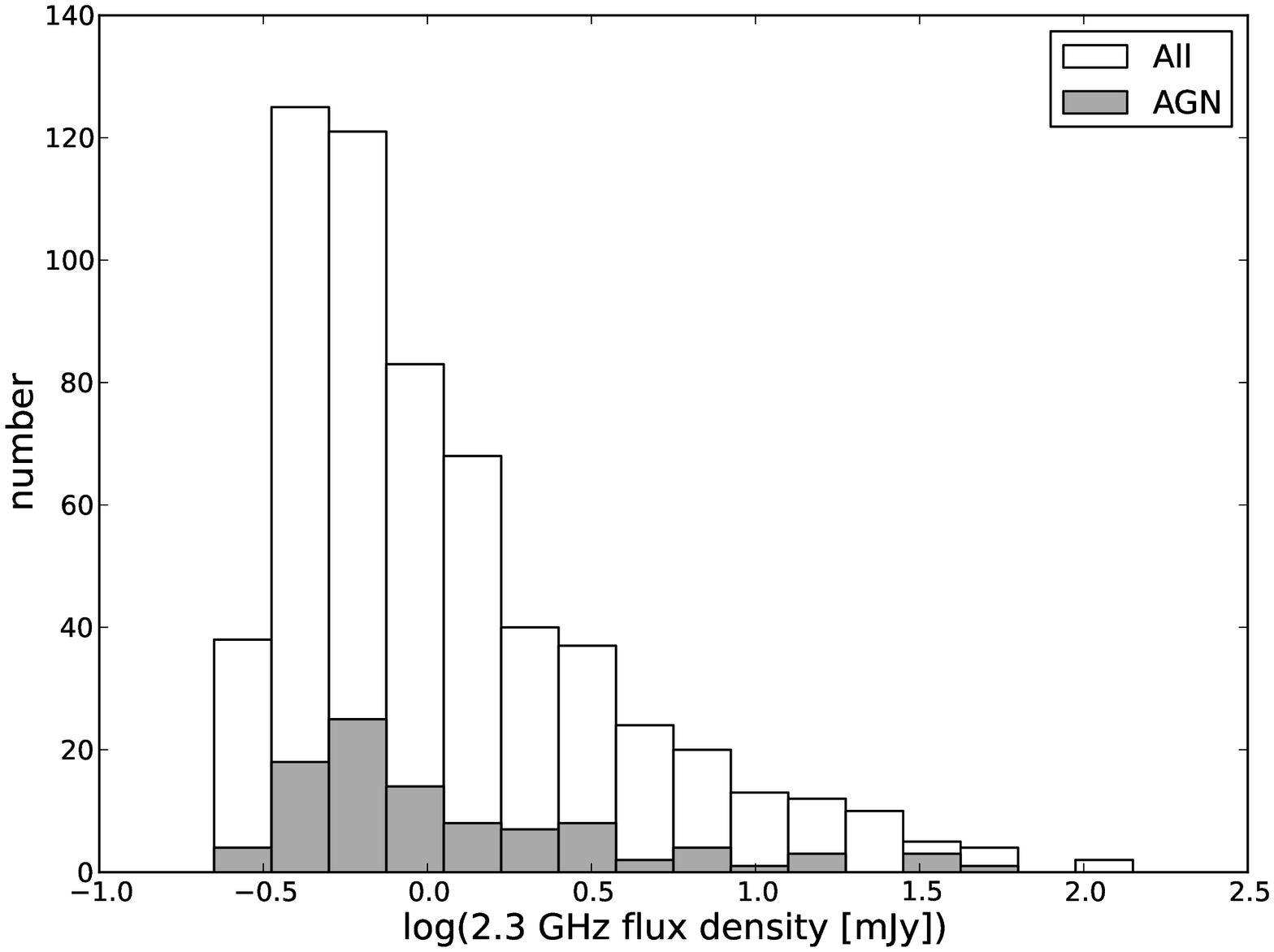}
\caption{Distribution of 1.4\,GHz (hi-res) and 2.3\,GHz flux densities
  in the final catalogue. AGN classifications of the sources are from
  \cite{Mao2012}.}
\label{fluxHist}
\end{figure*}

\begin{figure}
\centering
\includegraphics[width=0.5\textwidth]{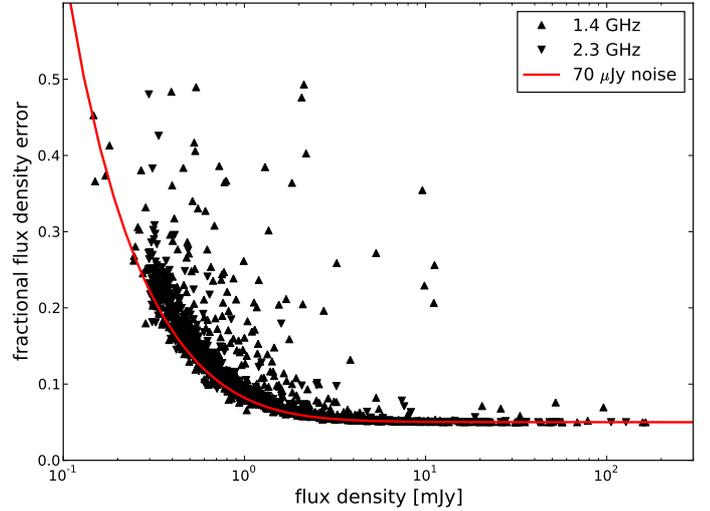}
\caption{The fractional flux density errors for all ELAIS-S1 and CDF-S
  sources as function of their flux density. The solid red line
  indictates how the fractional error should decrease with increasing
  flux density in an image with an overall rms noise of $70\,\mu$Jy
  according to Eqn.~\ref{rmsError}. A calibration error of 5\% is
  assumed.}
\label{fluxError}
\end{figure}

Histograms illustrating the distribution of flux densities in the
final catalogues are shown in Fig.~\ref{fluxHist}, the relative flux
density errors as a function of flux density are presented in
Fig.~\ref{fluxError}. Further spectral index properties are investigated in the
next Section.

\FloatBarrier

\section{Spectral index properties}
\label{spix}
\begin{figure}
\centering
\includegraphics[width=0.5\textwidth]{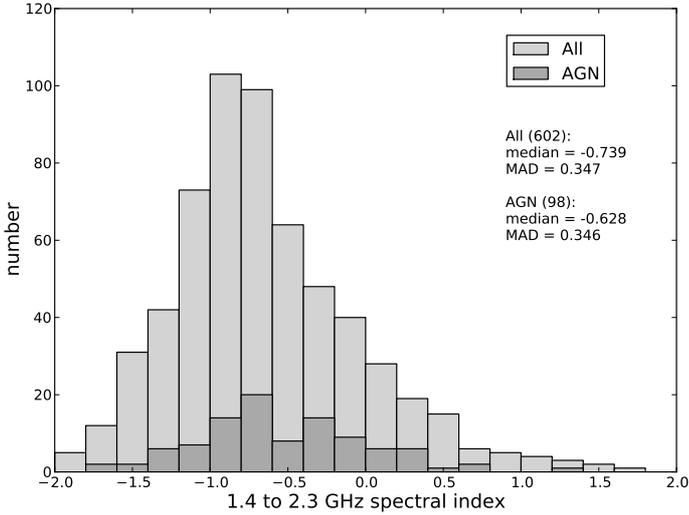}
\caption{Distribution of 1.4\,GHz to 2.3\,GHz spectral indices in the
  final catalogue for all sources { with unambiguously detected 2.3\,GHz counterpart and both the 1.4\,GHz and 2.3\,GHz flux densities exceeding 0.3\,mJy}. AGN classifications of the sources are from \cite{Mao2012}.}
\label{spixHist}
\end{figure}

The distribution of spectral indices { for our sample of 631 1.4\,GHz sources with unambiguously identified 2.3\,GHz counterpart} is shown in
Fig.~\ref{spixHist}.

\begin{figure}
\centering
\includegraphics[width=0.5\textwidth]{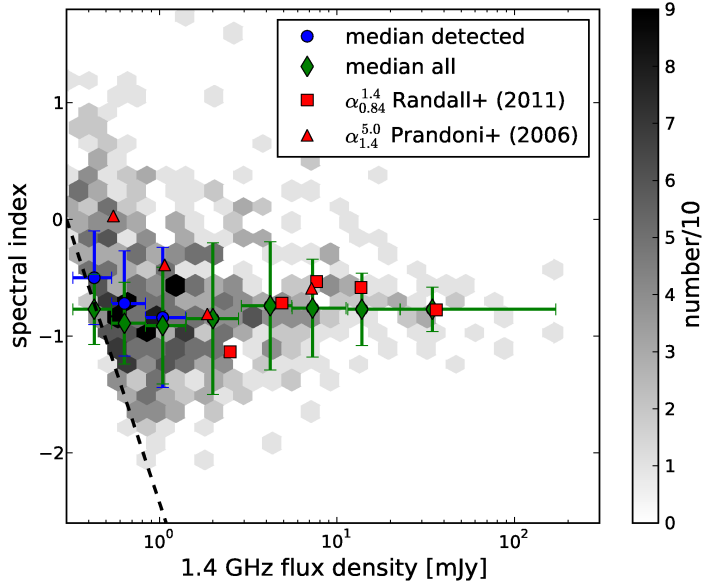}
\caption{Spectral index vs. flux density. { The gray 2D histogram shows our sample of 631 sources with 1.4\,GHz and 2.3\,GHz detections. Since the sensitivity of the 2.3\,GHz observations is $\sim3$ times lower than that of the 1.4\,GHz data, the black dashed line indicates the minimum spectral index that could be measured with our data. Therefore, the median spectral indices (blue circles, x-error bars indicate the flux density bin size, y-error bars correspond to the MAD) for just the 2.3\,GHz detected sample show significantly flatter values towards lower flux densities. When including also upper limits from the 2.3\,GHz non-detected sources, this trend completely vanishes (blue diamonds). Hence we conclude that there is no evidence for a flattening of the mean spectral index with lower flux density.}}
\label{spixFlux}
\end{figure}

With a median spectral index of $-0.74$ our sample matches the
canonical value for synchrotron radiation of $-0.7$ very
well. However, we note that the distribution is broad with a median
absolute deviation of 0.35. We here use the median absolute deviation
defined as
$\mathrm{MAD}\equiv\mathrm{median_i}(|x_{\mathrm{i}}-\mathrm{median_j}(x_{\mathrm{j}})|)$
as a measure for the spread of a distribution since it is more
robust against outliers than, for instance, the normal standard
deviation, $\sigma$. In case of a perfect Gaussian distribution,
$\sigma$ and $\mathrm{MAD}$ are related via
$\sigma\approx1.4826\,\mathrm{MAD}$. This also compares nicely to
other studies such as \cite{Prandoni2006} whose sample has
$\mathrm{MAD=0.33}$. The spread of the \cite{Ibar2009} sample is
slightly smaller with $\mathrm{MAD}=0.27$. This difference is due to
the larger number of sources with flat or inverted spectra in our
sample.
{ We repeated the analysis including upper limits \citep[following the survival analysis techniques developed in][]{Feigelson1985} for all 1.4\,GHz sources which do not have a 2.3\,GHz counterpart and are not affected by resolution effects (see Sect.~\ref{catalog}). These sources do have a 2.3\,GHz flux density less than three times the rms noise at the respective position. The median spectral index then becomes slightly steeper with $-0.81$ and a MAD of 0.38. The main reason for only having this slight change is the large difference between the 1.4\,GHz sensitivity (about 30\,$\mu$Jy) and the 2.3\,GHz sensitivity (70-80\,$\mu$Jy). The number of sources detected at both 1.4\,GHz and 2.3\,GHz is 631 whereas there are 552 sources with an unambiguous upper limit at 2.3\,GHz.
We note that the errors of our spectral indices become quite large for low flux densities. This is because currently available radio data and the flux densities extracted from them have typical uncertainties at the 10\% level. This is mostly due to either inhomogeneous data processing and calibration strategies as well as the widely adopted method of fitting 2-dimensional Gaussians to extract flux densities \citep[see e.g.][]{Condon1997}. Future large-scale continuum surveys, such as ASKAP/EMU, are attempting to reach flux density errors of only 1\% by developing both novel, fully automated calibration as well as source extraction techniques \citep{Norris2011}.

\subsection{Spectral index vs. flux density}
A plot of the spectral index of our sources against their flux density is shown in 
Fig.~\ref{spixFlux}. We point out that the sensitivities of our 1.4\,GHz and 2.3\,GHz data differ by a factor of about 3. Therefore, progressively negative spectral indices cannot be measured anymore towards fainter 1.4\,GHz fluxes (as indicated by the black dashed line in Fig.~\ref{spixFlux}). For instance, a source with $S_{\mathrm{1.4\,GHz}}=200\,\mu\mathrm{Jy}$ can only have a 2.3\,GHz counterpart when exhibiting a spectral index $\ga0.5$ since the $3\sigma$ detection level of our 2.3\,GHz data is about $250\,\mu\mathrm{Jy}$ (see Sect.~\ref{observations}). To account for this fact, we also included upper limits on spectral index for all 1.4\,GHz sources that unambiguously show no 2.3\,GHz counterpart (the upper limits were calculated using $S_{\mathrm{2.3\,GHz}}<3\sigma$ where $\sigma$ is the local rms noise for a point source) and are not affected by resolution issues (see Sect.~\ref{catalog}, e.g. a close double in 1.4\,GHz that would be inseparable at the 2.3\,GHz resolution). With these upper limits, we calculated median spectral indices for different flux density bins to investigate a potential flattening of the mean spectral index towards fainter flux densities. As expected, the median spectral indices for only the detected sample (so the 631 sources with 2.3\,GHz counterparts) show a clear trend of flatter spectral indices with lower flux density (Fig.~\ref{spixFlux}, blue circles). Including the upper limits into the median calculation, this trend vanishes immediately (Fig.~\ref{spixFlux}, blue diamonds): the lowest flux density bin (about 300-600\,$\mu$Jy) is completely dominated by the upper limits and can hence not be included in further analysis whereas the two flux density bins at 700\,$\mu$Jy and 1\,mJy are now a even a little steeper than the ones at higher flux densities. Hence, we find no evidence for a flattening of the spectral index
towards the lowest flux densities. This is consistent with the findings by \cite{Randall2012} who
used three frequencies (840\,MHz, 1.4\,GHz and 2.3\,GHz) for their
spectral index calculation and extend their findings to fainter flux
limits. \cite{Ibar2009} come to the same conclusion, too.}


{ A flattening of the spectral index with lower flux density levels, as e.g. seen by \cite{Zhang2003}, was initially suspected by \cite{Windhorst1993}. A major line of argument is that, at sub-mJy levels, a change in population takes place: At mJy levels, the radio sky is completely dominated by AGN \citep[see e.g.][]{Padovani2011}, so the dominant emission mechanism is non-thermal synchrotron radiation. An intrinsically flatter spectrum would e.g. be produced by dominating emission due to thermal bremsstrahlung, but this makes up only a marginal fraction even in starburst galaxies \citep[see][]{Condon1992}. In addition, recent work by several authors indicate that the transition from AGN-dominated to SF-dominated radio sources takes place below a flux density of 100\,$\mu$Jy. For instance, \cite{Padovani2011} argue that AGN still dominate the source population down to 100\,$\mu$Jy and \cite{Smolcic2008} claim that the fraction of SF-powered radio sources has a nearly constant value of about 40\% for flux levels between 50\,$\mu$Jy and 700\,$\mu$Jy. A similar result is obtained by \cite{Seymour2008} investigating the differential number counts in the 13 Hour XMM-{\it Newton}/{\it Chandra} Deep Field Survey. Hence, this change in the faint radio population cannot account for our findings since we do not reach such faint flux density levels.}

With our catalogue of 631 sources with spectral indices, we now focus
on the application of the spectral index in terms of its predictive
quality. Since it is a quantity which is relatively easy to measure,
the spectral index would be a useful tool to characterise sources in
large-scale surveys, if a dependence on other properties of interest
such as source type, redshift or other scaling relations can be
established.

\subsection{Spectral indices as object type discriminator}
\label{sec:object_type}
The radio spectral index is commonly regarded as a possible tool for
discriminating different radio source types, hence emission
mechanisms. For instance, \cite{Huynh2007} select AGN based { on} the
assumption that, at low frequencies, only AGN can produce flat or
inverted spectral indices because of self-absorption in an optically
thick medium. Hence they propose a cut between $\alpha=-0.3$ and
$\alpha=0$ to classify sources as AGN.

\begin{figure}
\centering
\includegraphics[width=0.5\textwidth]{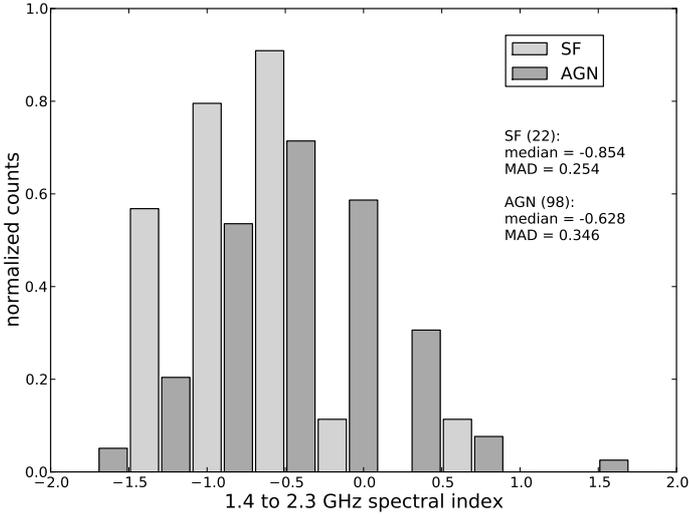}
\caption{Histogram illustrating the distribution of spectral indices of spectroscopically classified AGN and SF galaxies.}
\label{spixHistType}
\end{figure}

With our source sample, we can examine this classification
scheme. { We cross-matched all our sources with and without unambiguous 2.3\,GHz counterpart to} the spectral classifications for ATLAS sources by \cite{Mao2012}. { This cross-match lead to 132 resultant sources of which we can identify 22 SF galaxies and 98 AGN among our sample with measured spectral indices. The remainder 12 sources could either not be reliably classified by means of their optical spectrum (4 sources), or they only have spectral index upper limits (5 AGN, 3 SF galaxies). We suspect that the reason for most spectroscopically classified sources having actual spectral index measurements is due to the fact that the target sources for spectroscopy were chosen such that stronger radio sources were favored against weaker ones. Therefore, the spectroscopically classified sample shows a significantly higher mean flux density at 1.4\,GHz so that most of these sources are detected at 2.3\,GHz, too. Hence, since the vast majority of cross-identifications both have reliable spectral classifications and accurately measured spectral indices, we chose to omit the few sources that miss some information and proceed with these 120 sources}. A histogram showing the distribution of spectral indices among those 120 sources is presented in Fig.~\ref{spixHistType}.

As one can clearly see, the AGN population shows significantly flatter
spectral indices than the SF galaxies with median values of $-0.63$
for AGN and $-0.85$ for SF galaxies. Note that there are only
two SF galaxy exhibiting a spectral index greater than
$-0.5$. Therefore, flat or inverted spectral indices can reliably be
associated with dominant AGN activity. On the other hand, steep spectral indices (i.e. $\alpha<-1.0$) are not associated with any source type as the histogram clearly shows. Therefore, uniquely steep spectra cannot be used as discriminator for source type.

\begin{figure}
\centering
\includegraphics[width=0.5\textwidth]{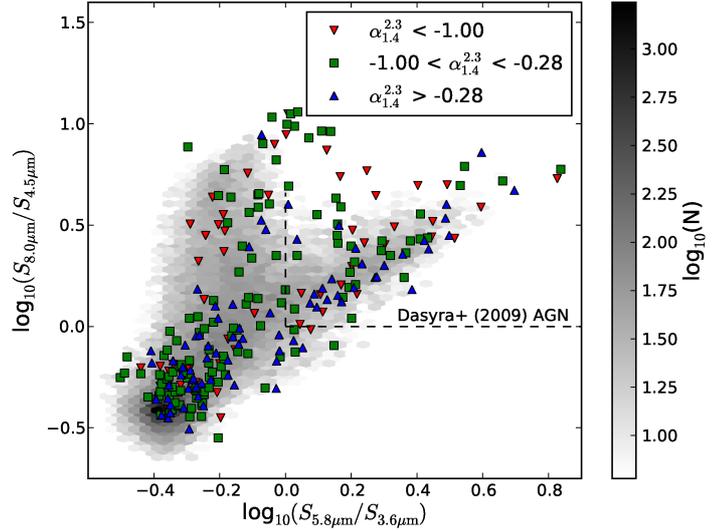}
\caption{``Lacy diagram'' of our source sample. For comparison, all
  sources detected in all four IRAC bands within the SWIRE survey of
  ELAIS-S1 and CDF-S are plotted as 2-dimensional histogram. The black
  dashed line indicates the locus of AGN in this diagram as proposed
  by \cite{Dasyra2009}.}
\label{lacy}
\end{figure}

Because the vast majority of (radio) sources detected in contemporary
and forthcoming surveys will not have any spectroscopic information,
we also searched for correlations between the spectral index and other
photometric object type indicators. A popular way of photometrically
selecting AGN without being biased by extinction is the use of an
infrared color-color diagram based on the four channels of the IRAC
instrument \citep{Fazio2004} aboard the {\it Spitzer} Space
Telescope. Originally proposed by \cite{Lacy2004} and therefore often
referred to as the ``Lacy diagram'', this plot has been further
investigated \citep[for an overview see][]{Soifer2008} so that there
are currently reliable template spectral energy distributions (SEDs)
available covering the entire optical to MIR regime
\citep[e.g.][]{Assef2010}. To construct a ``Lacy diagram'' for our
source sample, we cross-matched the radio sources with the SWIRE
catalogue, data release 3, to obtain IRAC flux densities. We found 265
secure identifications where all four IRAC fluxes are available. The
corresponding plot is presented in Fig.~\ref{lacy}.
For comparison, we also plotted all SWIRE-detected sources in the
CDF-S and the ELAIS-S1 field as color-coded, 2-dimensional
histogram. The locus of potential AGN in this diagram with respect to
various contaminating effects, predominantly line emission by
polycyclic aromatic Hydrocarbons (PAHs), was determined by
\cite{Dasyra2009} and is indicated in Fig.~\ref{lacy} with a dashed
line. We split our source sample in three sub-samples based on their
spectral index: steep spectrum sources are classified as having
$\alpha<\mathrm{median}(\alpha)-\mathrm{MAD}$, normal spectrum sources
satisfy
$\mathrm{median}(\alpha)-\mathrm{MAD}<\alpha<\mathrm{median}(\alpha)+\mathrm{MAD}$
and flat/inverted spectrum sources show
$\alpha>\mathrm{median}(\alpha)+\mathrm{MAD}$. One can see that flat or inverted spectrum sources are more frequent on the ``AGN branch'' of the diagram whereas the location of star-forming galaxies (the branch that goes straight up) is more or less avoided by these sources. This ``star formation branch''($\log(S_{\mathrm{5.8\,\mu
    m}}/S_{\mathrm{3.6\,\mu m}})<0$) is dominated by normal and steep spectrum
objects with only a very few flat/inverted spectrum sources.

\begin{figure}
\centering
\includegraphics[width=0.5\textwidth]{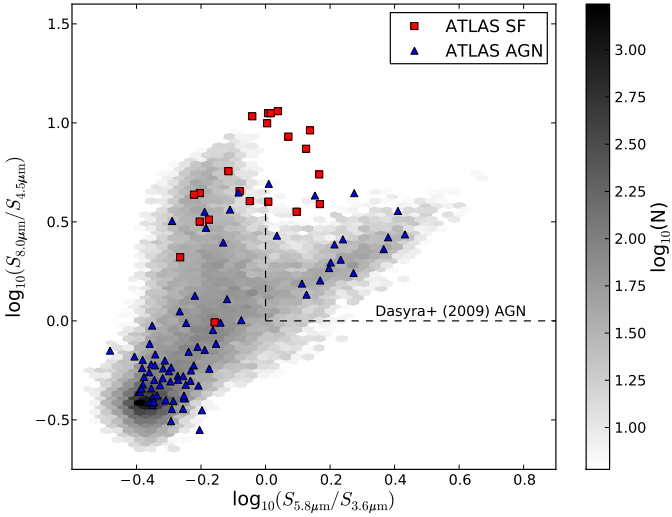}
\caption{``Lacy diagram'' of our source sample with spectroscopic classification from \cite{Mao2012}. As in Fig.~\ref{lacy}, all
  sources detected in all four IRAC bands within the SWIRE survey of
  ELAIS-S1 and CDF-S are plotted as 2-dimensional histogram. The black
  dashed line indicates the locus of AGN in this diagram as proposed
  by \cite{Dasyra2009}.}
\label{lacy2}
\end{figure}
To summarise, the radio spectral index can at least be used to
identify a fraction AGN, namely those with optically thick media
that cause their radio emission to flatten because of self-absorption
energy losses. However, spectral indices are not a strong
discriminator for the source type because (i) outliers from this
general trend are frequent and (ii) optically thin AGN cannot be
separated from SF galaxies with spectral indices. This is further highlighted in Fig.~\ref{lacy2} where we plotted the ``Lacy diagram'' for all ATLAS sources with spectroscopic classification \citep{Mao2012}. It becomes obvious that the ``AGN branch'' in this plot is really occupied by AGN only whereas the ``star formation branch'' both hosts SF galaxies and AGN. Hence, also a selection based on MIR color criteria leads to the result that a AGN can be reliably selected without contamination by star forming objects. The other way round, selecting SF galaxies only without AGN contamination, is not possible by means of MIR color selection, similar to a selection based on radio spectral indices.

\subsection{The $z$--$\alpha$ relation}
\label{z-alpha}

The $z$--$\alpha$ relation is an empirical relation seen between the
spectral index of radio sources in a flux-limited sample and their
redshifts. It states that higher-redshift radio sources tend to
exhibit steeper spectral indices, typically with $\alpha<-1$. It was
found by various authors using different large-scale radio surveys at
the Jy and mJy level, see e.g. \cite{Athreya1998},
\cite{DeBreuck2000,DeBreuck2001,DeBreuck2006}, \cite{Klamer2006},
\cite{Afonso2011}, and \cite{Ker2012}. The first quantitative
formulation of this relation was given by \cite{Verkhodanov2010} using
a large sample ($\sim2500$) of distant radio galaxies selected from
various surveys across the entire sky. Above a redshift of 1, they
find clear evidence for such a correlation and quantify it by a linear
fit $\alpha=(-0.70\pm0.02)+(-0.15\pm0.01)z$ out to $z\sim5$. The
currently favoured explanation for this effect is based on cosmic
expansion \citep{Klamer2006}: At higher redshifts, when the universe
was denser on average than at present, the radio jets of high-z AGN
had to penetrate into a denser intergalactic medium where they are
compressed, similar to the jets of radio galaxies located in the cores
of dense clusters in the nearby universe. This compression then leads
to the characteristic steepening of the spectral index due to
synchrotron and inverse-Compton energy losses of the relativistic
electrons.

\begin{figure}
\centering
\includegraphics[width=0.5\textwidth]{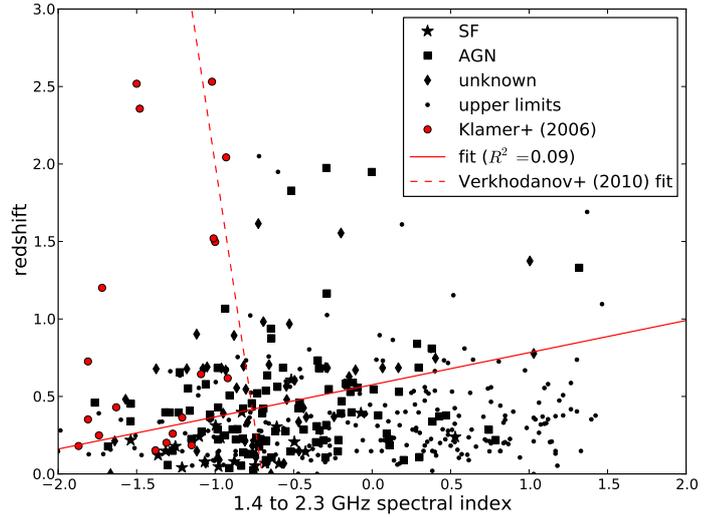}
\caption{Redshift versus spectral index plot for all sources with
  spectroscopic redshift in our sample. { Sources with an actual measurement of $\alpha$ are split in spectroscopic classes (AGN: black square, SF: black star, unknown: black diamond). Sources with an upper limit on $\alpha$ only (black dots) were not split up.} Note that $R^2=0.09$ for the
  fit to our data points (black symbols) indicates highly uncorrelated
  quantities. A slight correlation is usually assumed when $R^2>0.5$
  with good correlation for $R^2>0.85$.}
\label{spixZ}
\end{figure}

With the spectroscopic information from \cite{Mao2012}, we have
investigated the $z$--$\alpha$ relation in our sub-mJy sample. From
our initial 631 sources { with measured spectral index}, 169 (27\%) have spectroscopic redshifts. { In addition, we have 236 sources with spectroscopic redshift and an upper limit on $\alpha$ only.} A
plot of these redshifts versus the spectral index { or the upper limits, respectively,} is shown in
Fig.~\ref{spixZ}. { We again note that the 169 sources with spectroscopic redshifts and actual spectral index measurements are not affected by sensitivity bias (see Sect.~\ref{spix}) since they all have 1.4\,GHz flux densities in excess of 0.8\,mJy and hence spectral indices of at least -2.0 can be measured for them, given our 2.3\,GHz sensitivity.}

Surprisingly, in this diagram, there is no obvious correlation between
spectral index and redshift. Note that this result is also independent
of the spectroscopic classification of the objects. But since radio sources with significant redshift are, due to the sensitivity limits of currently available telescopes, all AGN-powered (even a galaxy forming stars at a very high rate of $1000\,M_{\odot}\,\rm{yr}^{-1}$ would be undetectable within ATLAS for any redshift $z>1.2$), the $z$--$\alpha$ relation is yet only investigated for AGN.

This finding (the absence of { a} $z$--$\alpha$ relation) contradicts the contemporary literature, and we offer two
potential explanations: First, our sample of sources is very small
(only 9 sources) for the redshift range $z>1$ were the
$z$--$\alpha$ relation becomes important
\citep[see][]{Verkhodanov2010}. Hence, our findings rely on small
number statistics and have therefore to be treated with care. Second,
there could be a change in the radio source population between our
sub-mJy sample and the samples used in other work which are mainly
much more luminous sources with flux densities of several tens to
hundreds of mJy (despite their significant redshifts, so these sources
are very luminous). We also point out that \cite{Ker2012} found only a slight
correlation between spectral index and redshift. They argue that the
relation becomes fully applicable only for resolved sources, if the
explanation is correct that the $z$--$\alpha$ relation arises from
high-redshift radio lobes working against the denser intergalactic
medium. This reasoning is supported by our findings because nearly all
of our sources are compact. In order to effectively select potential
high-z radio sources, \cite{Ker2012} suggest to use the relation
between $K$-band flux density (or also {\it Spitzer} 3.6\,$\mu$m IRAC
channel) and redshift which is much tighter than the traditional
$z$--$\alpha$ relation, see also \cite{Norris2011b}. This relation was
also investigated by \cite{Middelberg2011} to select a
certain class of objects which is suspected to be located at very high
redshifts ($z>2$), the so-called Infrared-Faint Radio Sources (IFRS).

\begin{figure}
\centering
\includegraphics[width=0.5\textwidth]{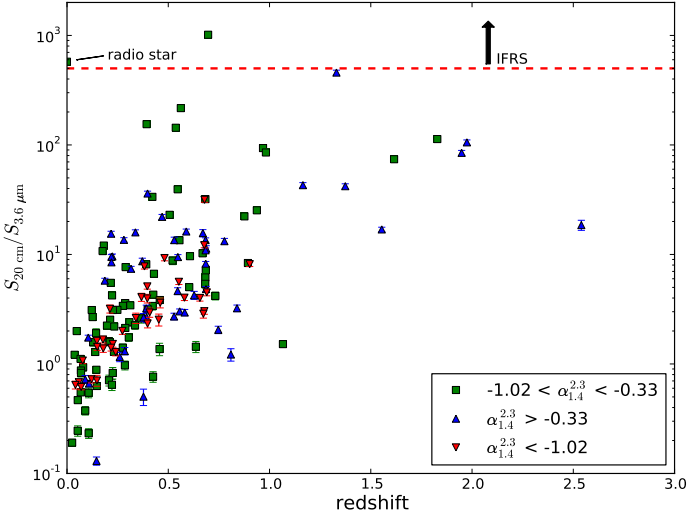}
\caption{$S_{\mathrm{20\,cm}}/S_{\mathrm{3.6\,\mu m}}$ flux ratios for our spectral index source sample. Note the clear but not very tight trend of increasing redshift with increasing $S_{\mathrm{20\,cm}}/S_{\mathrm{3.6\,\mu m}}$. The limiting value for IFRS is shown as red dashed line.}
\label{ifrs}
\end{figure}

According to \cite{Zinn2011}, a source is deemed { an} IFRS if its measured ratio between radio (20\,cm) and near-infrared (NIR, mostly {\it Spitzer} IRAC 3.6\,$\mu$m) flux density exceeds a cutoff-value of 500 and, at the same time, its NIR flux density is less than 30\,$\mu$Jy. Those extreme sources (often exhibiting no 3.6\,$\mu$m counterpart at all, even in ultra-deep or stacked NIR images with noise levels reaching the 100\,nJy level) are very difficult to investigate, in particular regarding their redshift, since optical spectroscopy can only be done for such IFRS that at least have moderately bright optical counterparts. Therefore, we only have very few sources in our sample with spectroscopic redshifts that meet at least the first IFRS criterion. A real IFRS meeting both criteria is not in our sample. A plot illustrating the $S_{\mathrm{20\,cm}}/S_{\mathrm{3.6\,\mu m}}$ flux ratios of our source sample with respect to redshift is shown in Fig.~\ref{ifrs}.
Contrary to the $z$--$\alpha$ relation in Fig.~\ref{spixZ}, there is a clear trend of increasing redshift with increasing $S_{\mathrm{20\,cm}}/S_{\mathrm{3.6\,\mu m}}$ flux ratio. Although the scatter in the relation is fairly large, in particular for high flux ratios, this relation seems to be a much better estimator for the redshift of a source than the classical $z$--$\alpha$ relation. This is in very good agreement with the findings of \cite{Ker2012} who also see only a marginal correlation of the spectral index with redshift but state that by far the most efficient selection criterion for high-$z$ sources in complete samples is based on $K$-band magnitude.

We conclude that our data do not confirm the $z$--$\alpha$ relation, at least
for faint radio sources. This may have several explanations: On the one hand, the popular belief that the $z$--$\alpha$ relation is caused by the denser intergalactic medium (IGM) at higher redshifts against which the radio jets of powerful sources have to work and thereby lose energy \citep{Klamer2006}, may not be applicable to faint sources since they are mostly compact. For instance, the simulations used by \cite{Raccanelli2011} predict that compact Fanaroff-Riley type I (FR I) sources are more frequent at any redshift greater than 1 by several orders of magnitude compared to extended FR II sources. Therefore, the overall radio population, at least at the sub-mJy level, should mainly consist of such compact FR I sources for which the \cite{Klamer2006} explanation is less suitable. Furthermore, the $z$--$\alpha$ relation may only be an apparent correlation. Nearly all radio spectra of the highest-$z$ ($z>4$) radio sources known to date are significantly curved or even peaked at observed-frame frequencies of $\la$500\,MHz \citep[, their Fig.~5]{Ker2012}, a behavior that is indicative for either young sources or sources with recent merger activity \citep[e.g.][]{ODea1998,Randall2011}. Hence, spectral indices calculated between 1.4\,GHz and another radio frequency commonly in use today, e.g. 610\,MHz or 5\,GHz, measure the high-frequency flank of such a peaked spectrum which yields increasingly steep spectral indices the more the radio spectrum is curved or peaked. Therefore, sources selected by a steep spectral index are predominantly of these two types (either young or merger). Because those sources are more frequent at higher redshifts when the universe was denser (more merger) and younger (more young sources), it only appears that a selection based on steep spectral indices yields high-redshift objects despite the real astrophysical phenomenon that is selected is completely different, namely young or merger sources.

\subsection{Spectral indices and the radio--IR correlation}

The radio--IR correlation is deemed to be one of the tightest
correlations in astrophysics. It states that the radio emission of a
galaxy is positively correlated with its mid- and far-infrared
emission due to recent star-formation activity. The IR emission in
this scenario is produced directly by the star-formation processes
whereas the radio emission comes from supernova remnants related to
the star-formation event. After { its} discovery in the late 1980s
\citep[e.g.][]{DeJong1985,Condon1991}, it was subjected to intensive
studies while the technology of infrared telescopes advanced
\citep{Appleton2004} and is nowadays proven to hold over a wide range
in both flux density \citep{Boyle2007,Garn2009} and redshift
\citep{Huynh2010,Mao2011}.

\begin{figure}
\centering
\includegraphics[width=0.5\textwidth]{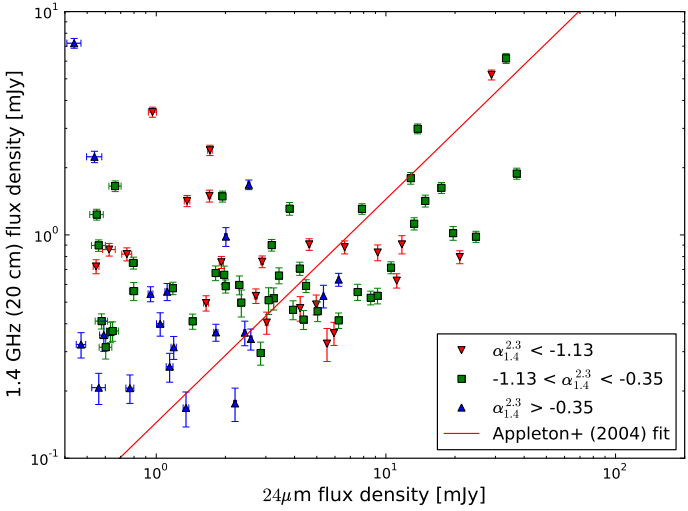}
\caption{Diagram illustrating the correlation between the flux
  densities at 24\,$\mu$m and 20\,cm (1.4\,GHz). The sources are
  divided in three spectral index bins according to
  $\alpha<\mathrm{median}(\alpha)-\mathrm{MAD}$,
  $\mathrm{median}(\alpha)-\mathrm{MAD}<\alpha<\mathrm{median}(\alpha)+\mathrm{MAD}$
  and $\alpha>\mathrm{median}(\alpha)+\mathrm{MAD}$ where MAD is the
  median absolute deviation of the spectral index of the entire
  24\,$\mu$m-detected sample.}
\label{24um20cm}
\end{figure}

We here use co-located {\it Spitzer} observations at 24\,$\mu$m from
the SWIRE survey \citep{Lonsdale2003} to investigate the dependence of
the correlation parameter, $q_{24}\equiv\log(S_{\mathrm{24\,\mu
    m}}/S_{\mathrm{20\,cm}})$ on the radio spectral index. Matching
our initial sample of 631 radio sources to the SWIRE IR catalogue with
a match radius of $5\arcsec$, 145 counterparts were
identified. Because the radio--IR correlation should only hold for
star-forming objects, we excluded all known AGN among these sources
based on the spectral classifications by \cite{Mao2012} and other AGN
indicators like morphology which was done in \cite{Norris2006} and
\cite{Middelberg2008}. After this rejection procedure, we were left
with a total of 86 sources with secure 24\,$\mu$m detection and no
signs of AGN activity. A plot showing $S_{\mathrm{24\,\mu m}}$ versus
$S_{\mathrm{20\,cm}}$ for these objects is presented in
Figure~\ref{24um20cm}.

The correlation in our source sample is , at least for their slope, well described by the fit of
\cite{Appleton2004} which was obtained using SWIRE data but over many
fields, not just ELAIS-S1 and CDF-S. The few sources that lie at the
top-left corner of the diagram, exhibiting an excess of radio
emission, are leftover AGN which were not removed from the sample
during the AGN rejection step.

\begin{figure}
\centering
\includegraphics[width=0.5\textwidth]{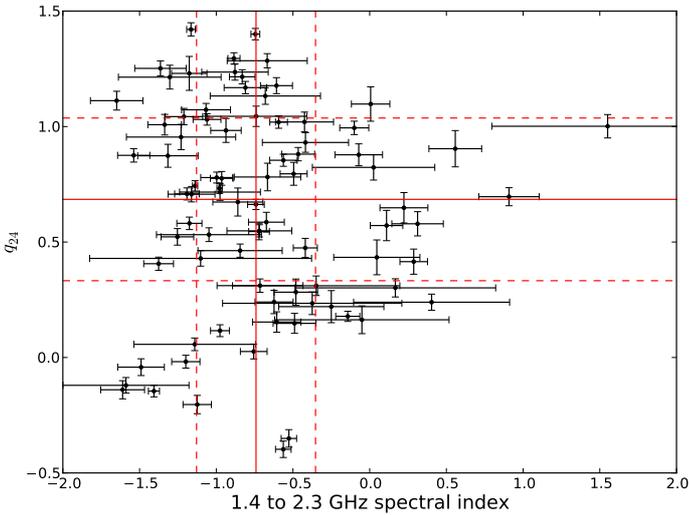}
\caption{$q_{24}$ values of our source sample plotted against spectral
  index. The solid red lines indicate the median of the two quantities
  while the dashed lines indicate the spread in terms of
  (median$\pm$MAD).}
\label{q24Spix}
\end{figure}

Calculating the correlation parameter $q_{24}$ for our sample, we find
a median value of 0.68 with a fairly large spread of
$\mathrm{MAD}=0.36$. When plotting the individual values of $q_{24}$
against the spectral indices of the sample (Fig.~\ref{q24Spix}), we
see no clear correlation between the two quantities.

Nevertheless, a general trend could be found that sources with a
spectral index steeper than the median value show a much larger spread
in $q_{24}$ values than sources with a spectral index flatter than the
median. This is increasingly surprising with the results of Sect.~\ref{sec:object_type} where we found that flat or inverted spectral indices are indicative for AGN-powered sources. Since AGN should not obey the radio--IR correlation, this tightening of the scatter in $q_{24}$ with increasing spectral index is still to be investigated. One possible explanation could arise from the findings of \cite{Padovani2011} who investigated star forming galaxies, radio-loud and radio quiet AGNs in the CDF-S. They found that regarding their luminosity function and its evolution, radio-quiet AGN are pretty much similar to star forming galaxies and not to their radio-loud siblings. Although we aimed to exclude AGN from the sample discussed in this section based on indicators such as radio morphology or spectroscopy, we predominantly eliminated radio-loud AGN and several radio-quiet AGN may still contaminating the sample (as suggested by the few flat and inverted spectrum sources in Fig.~\ref{q24Spix}). That these radio-quiet AGN do obey the radio--IR correlation is a natural consequence of the findings of \cite{Padovani2011} and already discussed in their work.

In summary, we conclude that the radio--IR correlation is not affected
by the radio spectral indices of sources in a flux-limited
sample. This is yet another proof for their remarkable universality which has been shown for redshift, flux density levels and now also for radio spectral properties.

\section{Summary and outlook}

We have conducted deep ($\sigma=70-80\,\mu$Jy) radio observations of
the ELAIS-S1 and CDF-S regions at 2.3\,GHz using the ATCA to supplement the sources detected at
1.4\,GHz within ATLAS with
spectral index information. Our final catalogue with sources well
detected at both frequencies and therefore with high quality spectral
indices comprise a total of 631 sources and is available in the online version of this article. The
median spectral index of our distribution is $-0.73$ with a median
absolute deviation (MAD) of 0.35. We have investigated the properties
of the population with respect to spectral index and tested correlations
between the spectral index and various other quantities. Our main
findings are as follows:

\begin{enumerate}

\item { We do not find evidence for a flattening of the spectral index towards fainter (several 100\,$\mu$Jy) flux densities. This is in good agreement with the majority of previous work by other authors \citep[e.g.][]{Randall2012,Ibar2009}. Since we also find that star-forming galaxies tend to have steeper radio spectra compared to AGN, we can support the current notion that star-forming galaxies become the dominant type of radio sources only at flux densities below 100\,$\mu$Jy since our sample does not reach such low flux densities.}

\item Considering the spectral index as discriminator for the object
  type, we find that flat- or inverted-spectrum sources are mostly AGN while SF galaxies show significantly steeper spectra with a
  median spectral index of $-0.85$ (AGN median spectral index is
  $-0.68$). This flattening is caused by self-absorption losses within
  optically thick AGN under physical conditions which are not in
  general met in SF galaxies. However, there is a significant number
  of outliers which affects the selection of AGN based on spectral
  indices. Furthermore, optically thin AGN exhibiting the canonical
  spectral index of $-0.7$ for synchrotron radiation can not be
  separated from SF galaxies. Therefore, we conclude that the spectral
  index is not a strong discriminator for the source type but can be
  helpful if there are no other, more sophisticated diagnostics.

\item We investigate the $z$--$\alpha$ relation which states that
  high-redshift radio galaxies show steeper spectral indices. The
  relation cannot be confirmed with our data. We suspect that this is explained either by the fact that the currently favored explanation for the $z$--$\alpha$ relation \citep{Klamer2006} is not applicable to the sub-mJy population or because of the complete failure of the $z$--$\alpha$ relation which may only be an apparent correlation. Since most high-z radio galaxies show curved or even peaked radio spectra, indicative for young or merger sources, a selection based on steep spectral indices may predominantly select those sources. But because both young and merger sources are more frequent at high redshifts when the universe was denser and younger, it appears to select high-$z$ sources.

\item We find no evidence for a dependence of the radio--IR
  correlation on radio spectral index. This therefore widens the
  universality of this correlation which is already known to hold over
  a wide range of flux densities and redshifts. However, we do see a
  trend for the scatter of $q_{24}$ which is larger for sources with
  steep spectral indices and becomes smaller while going to positive
  spectral indices.
\end{enumerate}

In conclusion, we want to point out that radio spectral indices can be
useful to characterise a radio-selected sample, but since there are
only rough correlations with other source properties, the predictive
power of spectral indices may have been overrated in the past. This is
unfortunate, in particular for future all-sky radio surveys such as
the Evolutionary Map of the Universe \citep[EMU,][]{Norris2011} to be
carried out with the Australia Square Kilometre Array Pathfinder
(ASKAP) telescopes in the southern sky or similar surveys in the
northern sky planned to be done with the Low Frequency Array
\citep[LOFAR,][]{Morganti2010} or the upgraded Westerbork Synthesis
Radio Telescope when equipped with the new focal plane array system
Apertif \citep{Rottgering2011}. A significant fraction of the sources
detected by these surveys at a typical noise level of 10\,$\mu$Jy will
have no other than radio data available, even considering future
surveys such as Pan-Starrs \citep{Kaiser2010}, SkyMapper
\citep{Keller2007}, and the Large Synoptic Survey Telescope
\citep{Ivezic2008} in the optical or the Wide-field Infrared Survey
Explorer \citep[WISE,][]{Wright2010} and ESO public surveys with the
VISTA and VST telescopes \citep{Arnaboldi2007} in the
infrared.

\begin{acknowledgements}
We thank our anonymous referee for many insightful comments and suggestions which improved the scientific validity of this paper a lot.\\
P.C.Z. acknowledges funding by the Ruhr-University Research School.
\end{acknowledgements}

\bibliographystyle{aa}
\bibliography{ATLAS_2.3GHz-catalog}

\end{document}